\documentclass[letterpaper,12pt]{article}
\usepackage[activeacute,spanish,french,english]{babel}
\usepackage[utf8]{inputenc}
\usepackage{amsfonts}
\usepackage{amsmath}
\usepackage{amssymb}
\usepackage{latexsym}
\usepackage{graphicx}% Include figure files
\usepackage{psfrag}
\usepackage{lscape}
\usepackage{multirow}
\usepackage{cite}
\usepackage{draftcopy}%
\usepackage{authblk}
\usepackage[active]{srcltx}

\newcommand{\al}{\alpha}
\title{CRITICAL CHARGE OF A SYSTEM WITH ONE ELECTRON AND FIVE OR SIX
  CHARGED CENTERS}

\author[1,2]{Héctor Medel Cobaxin\thanks{medel@nucleares.unam.mx}\thanks{hectormedel@itssat.edu.mx}} 
\affil[1]{Instituto de Ciencias Nucleares, Universidad Nacional
Aut\'onoma de M\'exico, Apartado Postal 70-543, 04510 M\'exico,
D.F., Mexico}
\affil[2]{Departamento de Investigaci\'on\\ Instituto Tecnol\'ogico Superior de San Andr\'es Tuxtla\\
  Carretera Costera del Golfo km 140+100, 95804 San Andr\'es Tuxtla, Mexico}

\date{}

\begin{document}

\maketitle

\begin{abstract}
 We consider a Coulomb system of one electron and five or six
  infinitely massive centers of charge $Z$: $(5Z,e)$ and
  $(6Z,e)$. Critical charges and the possible optimal geometrical
  configurations are found. It is shown that the domain of stability
  for $(5Z,e)$ is $0 < Z \leq Z_{cr}^{(5Z,e)}=0.350$ with the optimal
  geometrical configuration given by a dipyramid (equilateral triangle
  base) circumscribed in a prolate spheroid. For $(6Z,e)$ the
  stability is $0 < Z \leq Z_{cr}^{(6Z,e)}=0.335$ with the optimal
  geometrical configuration given by an octahedron (square base),
  circumscribed in an oblate spheroid. For both systems we obtain an
  indication that total energy at $Z=Z_{cr}$ has a square-root branch
  point singularity with exponent $3/2$.

\bigskip

% \noindent \footnotesize{{ Keywords}: One-electron molecular
%   ions; Critical charges; Variational method}

\end{abstract}

\section{Introduction}

In recent years it has been discovered that bound states for
one-electron systems with three/four protons exist in Nature in
presence of strong magnetic fields \cite{PR}. In particular, for the
molecular ion $H_3^{++}$ it has been shown that the geometrical
configuration changes depending on the field strength, being an
equilateral triangle for $10^8$ G$<B<10^{11}$ G and linear for
$10^{10}$ G $<B<4.414\times 10^{13}$ G. Systems with more than three
protons also exist in presence of such strong fields, but the only
linear configurations parallel to the field seems to be optimal.
Other other geometrical configuration might exist in such
circumstances but have not been studied, so far. In a very naive way
of thinking a clue to find other configurations is to assume the
proton to have a non-integer charge. Although the physics behind this
two phenomena are totally different, it is possible that for a given
field strength there will be a preferred direction for which a
geometrical configuration, different from linear, might be
realized. For example, in Ref.~\cite{TUR11:2411} it is presented the
case of one-electron systems with charged centers $(nZ,e)$, $n=2,3,4$,
turned out that there are bound states for positive charges less that
a certain critical one, $Z\leq Z_{cr}$, with non-linear optimal
geometrical configuration.

A restriction to integer values of charges do not appear in classical
electrodynamics and not even in more advanced theories such as the
theory of atomic-molecular physics, it is only in elementary particle
physics~\cite{schwinger,herdegen} where a justification for the
existence of integer charge appears. In classical electrostatics where
stable configurations of point charges are known to be absent
(Earnshaw's theorem), zero charge is a singular point where the nature
of interaction changes from repulsion to attraction. Usually, at a
singular charge the whole or some part of the potential vanishes. This
behavior is not exclusive of classical physics, it also can be found
in non-relativistic quantum electrodynamics where these singular
charges exist as well. However, a new phenomenon occurs - there are
some critical charges which separate the domain of the existence of
the bound states from the domain of non-existence, although the nature
of the potential remains unchanged. In some cases a system gets bound
at a critical charge with polynomialy-decaying eigenfunctions at large
distances, unlike standard exponentially-decaying eigenfunctions
\cite{Simon:77}. The well-known examples where such a transition, of
existence to non-existence, occurs are the square-well of finite
depth, the P\"oschl-Teller potential and the Yukawa potential. About
Coulomb molecular systems, there are some works related to molecular
ion $H_2^+$ with non-integer charge \cite{rebane,kais}. For more than
two charged centers and one electron, the first study was presented in
\cite{TUR11:2411} for $n=2,3,4$: $(2Z,e)$, $(3Z,e)$ and $(4Z,e)$,
where critical charges were calculated. The optimal geometric
configurations were found to correspond to Platonic solids; a line for
$(2Z,e)$, an equilateral triangle for $(3Z,e)$ and a regular
tetrahedron for $(4Z,e)$. For all studied systems, both atomic and
molecular, the total energy has a square-root branch point with
exponent $3/2$ at $Z=Z_{cr}$.

This work can be considered as a continuation of work
\cite{TUR11:2411} addressing five and six infinitely massive centers
of the same charge $Z$ and one electron. One of the goals is to find
the possible optimal geometrical configuration for which the system is
realized. Another goal is to identify the domain(s) in $Z$ where the
system is bound, focusing on the critical charges $Z_{cr}$ which
separate the domains of existence/non-existence of bound state.

This study is performed in the framework of non-relativistic quantum
mechanics. Atomic units $(\hbar=e=m_e=1)$ are used throughout, but
energies are given in Rydbergs.

\section{General considerations}

Let us consider a Coulomb molecular system which consists of $n$ fixed
charges $Z$ and one electron, $(nZ,e)$. The Hamiltonian which
describes this system as follows
\begin{equation}
  \label{Hn1}
  {\cal H}\ =\ -\frac{1}{2} \Delta \ +\ \sum_{i<j} \frac{Z^2}{R_{ij}}
  \ -\ \sum_{i=1}^{n}\frac{Z}{r_{i}}
  \ ,
\end{equation}
where $R_{ij}$ is the distance between charge centers $i$ and $j$, and
$r_{i}$ is the distance from the electron to $i$th charge center.

From a physical point of view, it is clear that such a system is not
bound, both for large positive and for negative $Z$ values. However,
it has to be bound at small $Z>0$. It is clear that there must be a
critical charge which separates the domains of existence and
non-existence of a bound state; one such a critical charge is $Z=0$,
the other one is at some finite $Z$, $Z=Z_{cr}$, which does not
represent any type of singularity of operator \eqref{Hn1}. It is worth
noting that it is possible to describe the same problem by considering
that the charged centers have positive charge equal to the unity and
the electron has negative charge greater than one.

In order to calculate the total energy $E(Z)$ we use the variational
method. We employ a physics-inspired trial functions. The choice of
the trial function is based on arguments of physical relevance,
i.e. the trial function should support the symmetries of the system,
has to reproduce the Coulomb singularities and the asymptotic behavior
at large distances adequately (for more details see
\cite{PR,turbinervar,turbinervar1,TL:2002}).

\section{System $(5Z,e)$}
Let us consider five fixed charges $Z$ and one electron. There is an
infinite number of configurations for this system, but only one that
is optimal. It is impossible to study all configurations but it is a
good guess to consider the most symmetric ones. Three particular
configurations were studied: pentagon, square pyramid and triangular
dipyramid. For the pentagon configuration, protons were located on the
the edges of the figure, defining the $x-y$ plane with origin in the
centroid, and $L$ the distance between two adjacent protons. In the
square pyramid configuration four protons were placed on the corners
of a square of side $L$, defining the $x-y$ plane with its sides
parallel to the axes; the fifth proton is placed on the $z-$axis
perpendicular to the square. Finally the triangular dipyramid
configuration is given by three centers forming an equilateral
triangle of side $d$ and two others which are symmetrically placed
perpendicular to the triangle coming through its center with distance
$2h$ between them.

% %\subsection{Trial Functions}

The variational method is used to obtain all numerical results. The
trial function is taken in a form of a symmetrized product of five
$1s$-Coulomb orbitals (Slater functions)
\begin{equation}
\label{trial5Ze}
\psi_g^{(5Z,e)} =\hat{S} e^{-\al_{1}r_1-\al_{2}r_2-\al_{3}r_3-\al_4r_4-\al_5r_5}\ ,
\end{equation}
where $\hat{S}$ is the symmetrizer with respect a permutation of the
charged centers. After doing permutations the trial function contains
120 terms. The function (\ref{trial5Ze}) depends on five $\al_i$
variational parameters. It is expected that this type of function
gives results with an accuracy of $\sim 10^{-3}$ in the total energy.

\begin{figure}[h!]
\centering
\psfrag{Z}{\huge{$Z$}}
\psfrag{Energy}{\huge{Energy (Ry)}}
\includegraphics[angle=-90, width=0.50\textwidth]{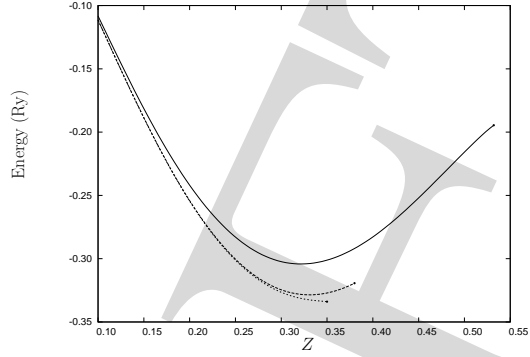}
\caption{\footnotesize{The total energy of systems $(5Z,e)$ as a function of the
  charge $Z$ for three different configuration: Pentagon (solid line),
  square pyramid (long-dashed line) and triangular dipyramid (dashed
  line). The solid line ends at $Z_{Pentagon}=Z^{(5Z,e)}_{cr}=0.532$,
  the long-dashed curve ends at $Z_{SqPyramid}=Z^{(5Z,e)}_{cr}=0.380$,
  and the dashed curve ends at $Z_{Dipyramid}=Z^{(5Z,e)}_{cr}=0.350$}}
\label{5ZeConfigs}
\end{figure}

\begin{figure}[htb]
\psfrag{Z}{\huge Z}
\psfrag{e}{\huge e}
\psfrag{d}{\huge d}
\psfrag{h}{\huge h}
\centering
\includegraphics[angle=00, width=0.20\textwidth]{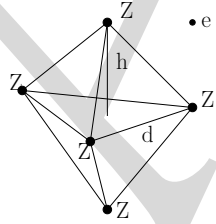}
\caption{\footnotesize{A system $(5Z,e)$: charges are located at vertices of the
  geometric figure. The electron is at point {\em{\bf e}}.}}
\label{5Ze}
\end{figure}

In order to find these critical charges, curves of the total energy as
function of the charge can be obtained for the three different
configurations, see Fig \ref{5ZeConfigs}. Calculations show that
critical charges are $Z_{cr}^{pentagon}=0.532$,
$Z_{cr}^{SqPyramid}=0.380$ and $Z_{cr}^{Dipyramid}=0.350$ for
pentagon, square pyramid and triangle dipyramid, respectively. The
curves also show that the most stable geometrical configuration is
given by the triangular dipyramid, see Fig.~\ref{5Ze}.

% realizes the optimal
% configuration (triangular dipyramid). It can be checked that this
% configuration is stable towards small deviations, see
%Figure.

%The energy dependence at equilibrium distances $d_{eq}$ and $h_{eq}$
%is a smooth function of the charge, see Fig. \ref{E:nz+e}.

For the optimal geometrical configuration, the triangular dipyramid,
it is possible to find the domains of metastability; i.e. points where
systems begin to have decay channels. These domains are given by
crossing points between one-electron and $n$ $Z$-charged systems,
$(nZ,e)$. Fig. \ref{E:nz+e} shows curves of the total energy as
function of the charge, $E(Z)$, for systems $(Z,e)$ (as reference
curve), $(3Z,e)$, $(4Z,e)$ (see \cite{TUR11:2411}), $(5Z,e)$ and
$(6Z,e)$, all taken in the optimal configuration. Thus, for charges
$Z\in (0.319,0.350)$ the system is metastable with two decay channels
\begin{eqnarray}
(5Z,e)&\to& (3Z,e)+Z+Z\nonumber \\
(5Z,e)&\to& (4Z,e)+Z
\end{eqnarray}
For $Z\in (0.24,0.319)$ the system is metastable with one decay
channel
\begin{equation}
(5Z,e)\to (4Z,e)+Z
\end{equation}
And finally for $Z<0.240$ the system is stable.

Beyond determining the critical charge, it is important to understand
the behavior of the energy, as function of the charge, close to
critically. In this way it turns out that this behavior is described by
means of a Puiseux expansion
\begin{equation}
\label{Pui}
    E(Z)= \sum_{n=0}^{\infty} a_n(Z_{cr}-Z)^{b_n}\ ,
\end{equation}
with the condition that $b_n<b_{n+1}$. The goal is to find parameters
$a_n$ and $b_n$ of this expansion.  Restricting the expansion
(\ref{Pui}) to a finite number of terms a fit of the total energy
calculated numerically. By making the fit it is found, in particular,
that exponents $b_n$ are very close to $m/2$ ($m=2,3,4,\ldots$); then,
it is convenient to assume $b_n=m/2$ and only determine the
coefficients $a_n$. The fit is based on data in the domain $0.30\leq Z
\leq 0.345$ (10 points). This behavior indicates that critical point
might be a square-root branch point.
\begin{eqnarray}
\label{5Zefitc}
E(Z)&=&-0.3339+0.1636(Z_{cr}-Z)\\\nonumber
&-&2.1606(Z_{cr}-Z)^{3/2}+15.1739(Z_{cr}-Z)^2\\\nonumber
&-&39.5356(Z_{cr}-Z)^{5/2} + \ldots \ ,
\end{eqnarray}
where the critical point is
\begin{equation}
\label{5Zecrv}
    Z_{cr}^{(5Z,e)}\ =\ 0.350 \ .
\end{equation}

For the critical point $Z=0$ (which is the singular point of the
Schr\"odinger equation), we study the behavior of the energy as
function of the charge. The behavior on the energy ($0\leq Z\leq 0.15$
with 15 points) is given by a Taylor expansion:
 \begin{equation}
\label{5Zefit0}
 E(Z)= -19.9541Z^2+109.699Z^3-207.039Z^4+\ldots
 \end{equation}
 Such a behavior does not provide an indication of a singular nature
 at $Z=0$. However, the total energy cannot be analytically continued
 to $\mbox{Re} Z <0$.

 Table \ref{fitcorr5Ze} shows a comparison between the results of the
 energy fit \eqref{5Zefitc} and fit \eqref{5Zefit0} near critical charges
 $Z=Z_{cr}$, $Z=0$, correspondingly, and data.
\begin{table}[h!]
\centering
\begin{tabular}{c c c}
\hline
 $Z$&$E_T$&Fit\\\hline
 0.10&\hspace{0.5cm}    -0.1113723&\hspace{0.5cm}     -0.111545 \\
 0.11&\hspace{0.5cm}    -0.1271357&\hspace{0.5cm}     -0.127211 \\
 0.12&\hspace{0.5cm}    -0.1428512&\hspace{0.5cm}     -0.142783 \\
 0.13&\hspace{0.5cm}    -0.1583834&\hspace{0.5cm}     -0.158202 \\
 0.14&\hspace{0.5cm}    -0.1736134&\hspace{0.5cm}     -0.173462 \\
 0.15&\hspace{0.5cm}    -0.1884371&\hspace{0.5cm}     -0.188607 \\\hline

 0.30&\hspace{0.5cm}    -0.3265247&\hspace{0.5cm}     -0.326525 \\
 0.31&\hspace{0.5cm}    -0.3291843&\hspace{0.5cm}     -0.329184 \\
 0.32&\hspace{0.5cm}    -0.3311328&\hspace{0.5cm}     -0.331133 \\
 0.33&\hspace{0.5cm}    -0.3324624&\hspace{0.5cm}     -0.332462 \\
 0.34&\hspace{0.5cm}    -0.3332823&\hspace{0.5cm}     -0.333282 \\
 0.35&\hspace{0.5cm}    -0.3339403&\hspace{0.5cm}     -0.333940 \\
\end{tabular}
\caption{\footnotesize{Total energy of $(5Z,e)$, obtained with (\ref{trial5Ze}), compared with the result of the fit (\ref{5Zefitc}) (for $0.30\leq Z\leq 0.345$) and (\ref{5Zefit0}) (for $0\leq Z\leq 0.15$).}}
\label{fitcorr5Ze}
\end{table}

\begin{figure}[h!]
\centering
\psfrag{R  h}{\huge{$h$ and $R$ }}
\psfrag{Charge}{\huge{$Z$}}
\includegraphics[angle=-90, width=0.5\textwidth]{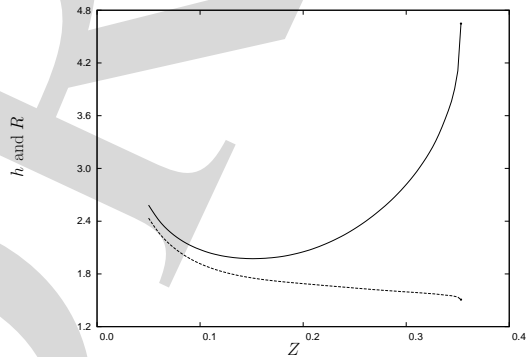}
\caption{\footnotesize{Equilibrium geometrical configuration: Height of charges $h$
  (solid line) and radius $R$ of circumscribed circle for charges
  fixed on vertices of an equilateral triangle (dashed line), all as
  function of the charge $Z$.}}
\label{5Zedistances}
\end{figure}

\begin{figure}[h!]
\psfrag{R}[bl][l][1][-90]{\huge R}
\psfrag{h}[bl][l][1][-90]{\huge h}
\centering
\includegraphics[angle=90, width=0.20\textwidth]{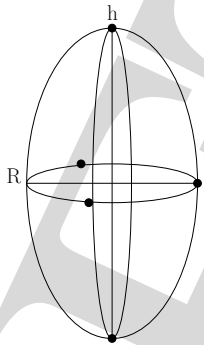}
\caption{\footnotesize{Equilibrium geometrical configuration for the system
  $(5Z,e)$, $Z<Z_{cr}$. Charges are situated on surface of prolate
  spheroid of semi-axes $R$ and $h$.}}
\label{elipse5Ze}
\end{figure}

Finally, Fig. \ref{5Zedistances} shows that the optimal geometrical
configuration is a triangular dipyramid with equilateral triangular
base. Its height $h$ is always greater than the radius of maximal
circular section of spheroid (circumscribed circle for the base)
$R$. This tells us that the charges are situated on a prolate spheroid
with semi-axis $R$, see Fig.\ref{elipse5Ze}. The form of the spheroid
changes with a charge variation, always remaining prolate.

\section{System $(6Z,e)$}
Now let us consider six fixed charges $Z$ and one electron. As in the
five-center case, in this system there are infinitely many geometrical
configurations. Among them there must exist one which is optimal. It
is possible to consider some of them, the most symmetric ones. Four
configurations are considered: $1)$ hexagon, the charges are placed on
the vertices of the figure of side $L$ lying on the $x\!-\!y$ plane
with the center in the origin of the plane and two charges on the
$x$-axis; $2)$ a pentagon pyramid, with five charges on the vertices
of a pentagonal base of side $L$ and one charge at height $h$ on the
perpendicular passing through the center; $3)$ an equilateral triangle
dipyramid with three charges on the vertices of the triangle of side
$L$, one charge in the center and other two on the perpendicular
passing through the center of the triangle with distance $2h$ between
them; and finally $4)$ a square dipyramid with four centers forming a
square of side $l$, and two other lying on the perpendicular, passing
through the center of the square, separated by $2h$ (octahedron).

The trial function for this case is taken in a form of a symmetrized
product of six $1s$-Coulomb orbitals (Slater functions)
\begin{equation}
\label{trial6Ze}
\psi_g^{(6Z,e)} =\hat{S} e^{-\al_{1}r_1-\al_{2}r_2-\al_{3}r_3-\al_4r_4-\al_5r_5-\al_6r_6}\ ,
\end{equation}
where $\hat{S}$ is the symmetrizer with respect a permutation of
charged centers. After performing permutations, the trial function
\eqref{trial6Ze} contains $720$ terms, and it depends on six $\alpha$
variational parameters, given an accuracy of $\sim 10^{-3}$ for the
total energy.

Because of the complexity of the calculations and the loss of accuracy
for small values of the charge, an energy curve cannot be fully
calculated. Instead the total energy is computed for some given values
of the charge. Table \ref{6ZeConfigsT} shows the total energy of the
different configurations for some fixed values of the charge $Z$.
\begin{table}[h]
\centering
\begin{tabular}{ccccc}
  \hline
  \multirow{2}{*}{Z}&\multicolumn{4}{c}{Energy ($Ry$)}\\\cline{2-5}
  &SD&TD&PP&Hexagon\\\hline
$0.300$&$-0.2855$&$-0.2821$&$-0.2539$&$-0.2373$\\
$0.335$&$-0.2749$&--       &$-0.2348$&$-0.2125$\\\hline
\end{tabular}
\caption{\footnotesize{The total energy as a function of the charge $Z$ for various configurations of the  $(6Z,e)$ system: square dipyramid (SD), triangular dipyramid (TD), pentagon pyramid (PP) and hexagon. For $Z=0.335$ the triangular dipyramid does not exist.}}
\label{6ZeConfigsT}
\end{table}

From Table \ref{6ZeConfigsT} we see that among the configurations
studied, the square dipyramid presents the lowest total energy, and we
conjectured that this configuration is optimal for the $(6Z,e)$
system.

\begin{figure}[h!]
\psfrag{Z}{\huge Z}
\psfrag{e}{\huge e}
\psfrag{l}{\huge l}
\psfrag{h}{\huge h}
\centering
\includegraphics[angle=00, width=0.25\textwidth]{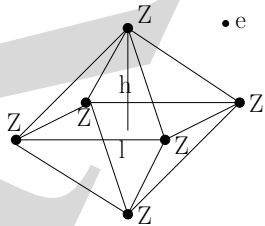}
\caption{\footnotesize{The system $(6Z,e)$. Charges are located at the vertices of
  the octahedron. The electron is at point {\em{\bf e}}.}}
\label{6Zegc}
\end{figure}

\begin{figure}[h]
\centering
\psfrag{Charge}{\huge{$Z$}}
\psfrag{Energy}{\huge{Energy (Ry)}}
\psfrag{(Z,e)}{\huge{$(Z,e)$}}
\psfrag{(3Z,e)}{\huge{$(3Z,e)$}}
\psfrag{(4Z,e)}{\huge{$(4Z,e)$}}
\psfrag{(5Z,e)}{\huge{$(5Z,e)$}}
\psfrag{(6Z,e)}{\huge{$(6Z,e)$}}
\includegraphics[angle=-90, width=0.5\textwidth]{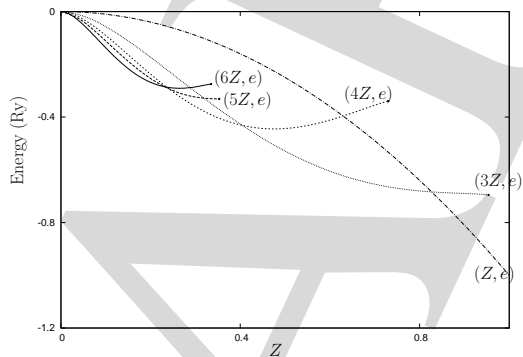}
\caption{\footnotesize{The total energy of systems $(Z,e)$ (dot-dashed line),
  $(3Z,e)$ (dotted line), $(4Z,e)$ (dashed line), $(5Z,e)$
  (long-dashed) and $(6Z,e)$ (solid line) as functions of the charge
  $Z$, all for optimal geometrical configuration. The dotted curve
  ends at $Z=Z^{(3Z,e)}_{cr}=0.9537$.  The dashed curve ends at
  $Z=Z^{(4Z,e)}_{cr}=0.736$. The long-dashed curve ends at
  $Z=Z^{(5Z,e)}_{cr}=0.350$. The solid curve ends at
  $Z=Z^{(6Z,e)}_{cr}=0.335$.}}
\label{E:nz+e}
\end{figure}

For the optimal configuration, see Fig.~\ref{6Zegc}, the critical
charges which separate the domains of existence and non-existence of
the bound state are at $Z=0$ and $Z_{cr}=0.335$. Thus, the system
$(6Z,e)$ can exist for charges $0<Z<Z_{cr}$. Fig.~\ref{E:nz+e} shows
the total energy as a function of the charge at equilibrium distances
$l_{eq}$ and $h_{eq}$. From Fig. \ref{E:nz+e} we can see the crossing
points with other curves, that specify the domains of metastability of
the system.

The domains of metastability are given in the following way:
for charges $Z\in (0.2879,0.3350)$ the system is metastable with three
decay channels
\begin{eqnarray}
(6Z,e)&\to& (3Z,e)+Z+Z+Z\nonumber \\
(6Z,e)&\to& (4Z,e)+Z+Z\nonumber\\
(6Z,e)&\to& (5Z,e)+Z
\end{eqnarray}
For $Z\in(0.2358,0.2879)$ the system is metastable with two decay
channel
\begin{eqnarray}
(6Z,e)&\to& (4Z,e)+Z+Z\nonumber\\
(6Z,e)&\to& (5Z,e)+Z
\end{eqnarray}
For $Z\in(0.2318,0.2358)$ the system is metastable with the single
decay channel
\begin{equation}
(6Z,e)\to (5Z,e)+Z\ ,
\end{equation}
and, finally, for $Z<0.2318$ the system is stable.

The behavior of the energy as a function of the charge close to
critical charge $Z_{cr}$, is given by the following Puiseux expansion
%(See Fig. \ref{fitz3++1}):
\[
E(Z)\ =\ -0.2749-0.0763(Z_{cr}-Z)-3.7767(Z_{cr}-Z)^{3/2}
\]
\begin{equation}
\label{6Zefitc}
+\ 22.813(Z_{cr}-Z)^2 - 54.9039(Z_{cr}-Z)^{5/2}
\end{equation}
where the critical point is
\begin{equation}
\label{6Zecrv}
    Z_{cr}^{(6Z,e)}\ =\ 0.335 \ .
\end{equation}
The result of the fit is based on data from the domain $0.32\leq Z\leq
0.33$ (20 points). This behavior indicates that the critical point
might be a square-root branch point with exponent 3/2.

Near the critical point $Z=0$ the behavior of the energy ($0\leq Z\leq
0.15$ with 15 points) is given by the Taylor expansion
 \begin{equation}
\label{6Zefit0}
 E(Z)= -28.1767Z^2+184.321Z^3-405.044Z^4+\ldots
 \end{equation}
 Such a behavior does not indicate a singularity in the point
 $Z=0$. However, the total energy can not be analytically continued to
 $\mbox{Re} Z <0$.

 The comparison between the fits of the energy (\ref{6Zefitc}) and
 (\ref{6Zefit0}) near critical charges $Z=Z_{cr}$, $Z=0$, and the data
 are shown in Table \ref{fitcorr6Ze}.
\begin{table}[htb]
\centering
\begin{tabular}{c c c}
\hline
 $Z$&$E_T$&Fit\\\hline
 0.1000 &\hspace{0.5cm}  -0.1376189 &\hspace{0.5cm}  -0.13795   \\
 0.1100 &\hspace{0.5cm}  -0.1549053 &\hspace{0.5cm}  -0.15491   \\
 0.1200 &\hspace{0.5cm}  -0.1715792 &\hspace{0.5cm}  -0.17123   \\
 0.1300 &\hspace{0.5cm}  -0.1874736 &\hspace{0.5cm}  -0.18692   \\
 0.1400 &\hspace{0.5cm}  -0.2024490 &\hspace{0.5cm}  -0.20209   \\
 0.1500 &\hspace{0.5cm}  -0.2163915 &\hspace{0.5cm}  -0.21639   \\\hline
 0.320&\hspace{0.5cm}    -0.2793243&\hspace{0.5cm}  -0.279328    \\
 0.322&\hspace{0.5cm}    -0.2786595&\hspace{0.5cm}  -0.278657    \\
 0.324&\hspace{0.5cm}    -0.2779993&\hspace{0.5cm}  -0.277997    \\
 0.326&\hspace{0.5cm}    -0.2773488&\hspace{0.5cm}  -0.277350    \\
 0.328&\hspace{0.5cm}    -0.2767147&\hspace{0.5cm}  -0.276718    \\
 0.330&\hspace{0.5cm}    -0.2761066&\hspace{0.5cm}  -0.276108    \\
\end{tabular}
\caption{\footnotesize{The total energy of $(6Z,e)$, obtained from (\ref{trial6Ze}), compared with the result of the fit (\ref{6Zefitc}) (for $0.320\leq Z\leq 0.330$) and (\ref{6Zefit0}) (for $0\leq Z\leq 0.15$).}}
\label{fitcorr6Ze}
\end{table}

\begin{figure}[h!]
\centering
\psfrag{R h}{\huge{$h$ and $R$ }}
\psfrag{Charge}{\huge{$Z$}}
\includegraphics[angle=-90, width=0.5\textwidth]{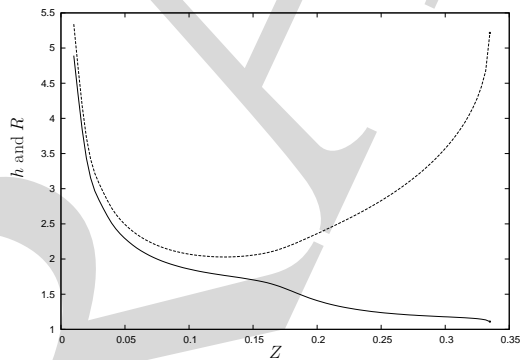}
\caption{\footnotesize{Equilibrium geometrical configuration: the height $2h$ is the
  distance between anti-polar charges (solid line) and radius $R$ of
  circumscribed circle for charges fixed on vertexes of the square
  (dashed line), all as function of the charge $Z$.}}
\label{6Zedistances}
\end{figure}

Finally, Fig. \ref{6Zedistances} presents curves for the distances
(height $h$, and radius of circumscribed circle $R$) as functions of
the charge $Z$ for the geometrical configuration of the system
(octahedron).

\begin{figure}[htb!]
\psfrag{h}{\huge h}
\psfrag{R}{\huge R}
\centering
\includegraphics[angle=00, width=0.30\textwidth]{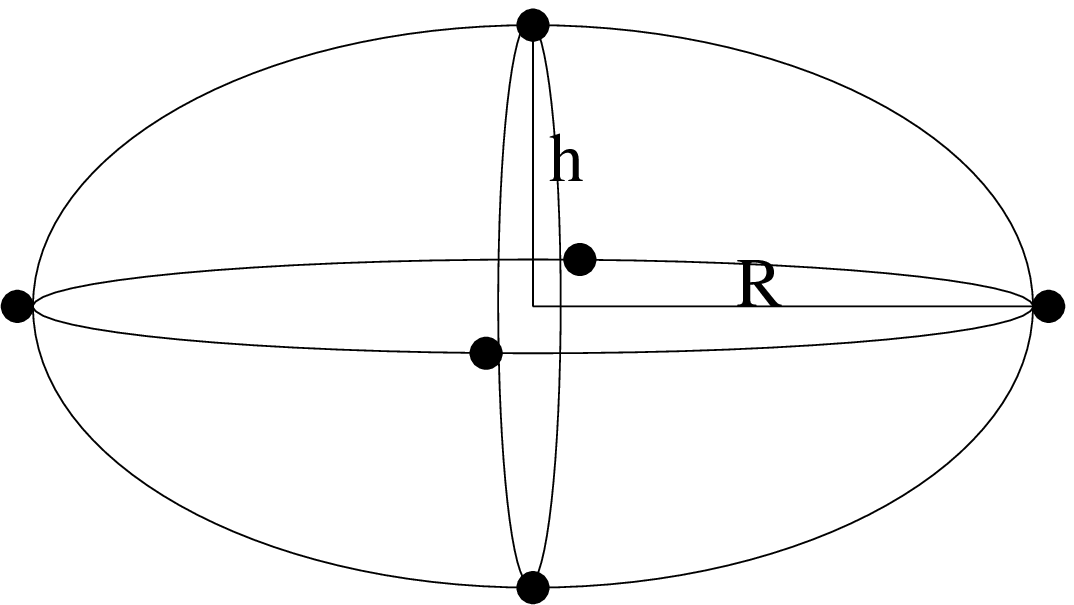}
\caption{\footnotesize{Equilibrium geometrical configuration for system $(6Z,e)$,
  $Z<Z_{cr}$: Charges are situated on the surface of an oblate
  spheroid of semi-axes $R$ and $h$.}}
\label{elipse6Ze}
\end{figure}

From Fig. \ref{6Zedistances} we see that the optimal geometrical
configuration is an octahedron, where the height $h$ is always smaller
than the radius of circumscribed circle $R$. This tell us that the
charges are placed on a oblate spheroid with semi-axes $R$, see
Fig.\ref{elipse6Ze}.
%%%%%%%%%%%%%%%%%%%%%%%%%%%%%%%%%%%%%%%%

%%%%%%%%%%%%%%%%%%%%%%%%%%%%%%%%%%%%%%%%%%%%%%%%%%%%%%%%%%%%%%%%%%%%%
\section{Conclusions}
In this paper we calculated for the first time the critical charges
for two molecular systems: $(5Z,e)$ and $ (6Z,e)$. For all those
systems the total energy and equilibrium distances vs. $Z$ are smooth
curves without any indication of charge quantization. Moreover the
optimal geometric configuration seems to be always the maximally
symmetric: a triangular dipyramid for $(5Z,e)$ and an octahedron for
$(6Z,e)$.

It is important to mention that the behavior near the critical charge
of the total energy as a function of the charge, $E(Z)$, indicates a
square-root branch point with exponent 3/2. This agrees with the previous
results found for various atomic and molecular systems
\cite{TUR11:2411}.

%\begin{acknowledgments}
\section*{Acknowledgements}
  The author is grateful to A. V. Turbiner, A. Alijah and
  W. Bietenholz for their valuable suggestions and useful comments,
  and for reading the manuscript. He further thanks J. C. L\'opez
  Vieyra for his assistance with computer calculations and interest in
  the present work. This work was supported in part by CONACyT grants
  {\bf 58962-F}, {\bf 171494} and {\bf 202139} (Mexico), and by the
  Computer Center of the Universit\'e de Reims Champagne-Ardenne
  (France). 
%\end{acknowledgments}

\end{document}